\documentclass[%
 reprint,
superscriptaddress,
 amsmath,amssymb,
 aps,
]{revtex4-2}

\usepackage{graphicx}
\usepackage{dcolumn}
\usepackage{bm}
\usepackage{verbatim}
\usepackage{hyperref}
\usepackage{braket}
\usepackage{xfrac}
\usepackage[usenames]{color}
\usepackage{bm}
\usepackage{soul}
\usepackage{subfigure}

\begin{document}

\preprint{Parameter Estimation of Gravitational Waves with a Quantum Metropolis Algorithm}

\title{Parameter Estimation of Gravitational Waves with a Quantum Metropolis Algorithm}

\author{Gabriel Escrig}
\email{gescrig@ucm.es}
\affiliation{Departamento de Física Teórica, Universidad Complutense de Madrid.}

\author{Roberto Campos}
\email{robecamp@ucm.es}
\affiliation{Departamento de Física Teórica, Universidad Complutense de Madrid.}
\affiliation{Quasar Science Resources, SL.}

\author{Pablo A. M. Casares}
\email{pabloamo@ucm.es}
\affiliation{Departamento de Física Teórica, Universidad Complutense de Madrid.}

\author{M. A. Martin-Delgado}
\email{mardel@ucm.es}
\affiliation{Departamento de Física Teórica, Universidad Complutense de Madrid.}
\affiliation{CCS-Center for Computational Simulation, Universidad Politécnica de Madrid.}


\begin{abstract}
After the first detection of a gravitational wave in 2015, the number of successes achieved by this innovative way of looking through the universe has not stopped growing. However, the current techniques for analyzing this type of events present a serious bottleneck due to the high computational power they require. In this article we explore how recent techniques based on quantum algorithms could surpass this obstacle. For this purpose, we propose a quantization of the classical algorithms used in the literature for the inference of gravitational wave parameters based on the well-known Quantum Walks technique applied to a Metropolis-Hastings algorithm. Finally, we develop a quantum environment on classical hardware, implementing a metric to compare quantum versus classical algorithms in a fair way. We further test all these developments in the real inference of several sets of parameters of all the events of the first detection period GWTC-1 and we find a polynomial advantage in the quantum algorithms, thus setting a first starting point for future algorithms.
\end{abstract}

\maketitle


\section{Introduction}
After 100 years of its prediction and an international effort spanning several decades, on September 14th, 2015, the two Advanced LIGO detectors simultaneously detected the first signal of a gravitational wave (GW)~\cite{primeradet}. This made it possible to open a new window through which to look at the universe, complementary to the observation of the electromagnetic spectrum. This new method promises the observation of astrophysical events never seen before, but its processing is costly. The computational bottleneck arises in the Bayesian inference of the parameters of the sources that generated the gravitational radiation and will grow when a larger number of parameters, higher simulation accuracy and more events to analyze are obtained due to the improvement of the detectors between observation periods.

Historically, the computational advances have been very large. Over the last 20 years, different types of algorithms have been proposed (see Refs~\cite{tiposinferencia, papertoni}), but those based on the formalism of Bayesian inference have stood out from the rest, making Bayesian parameter estimation the language of gravitational-wave astronomy. Among these algorithms the most important are the Metropolis-Hastings algorithms based on Monte Carlo Markov chains \cite{tesisphd}. In addition, a large number of software libraries have been developed (see Refs \cite{astropy, pycbc, bilby}), making it clear that the field of GW is growing.

Currently, the computational analysis cost is very large and will get larger for several reasons. First, in the near future, and with the continuous improvement of the detectors, many more black hole and neutron star mergers will be measured. For instance, in the next observing period O4, the LIGO-Virgo-KAGRA collaboration is expected to observe about $79^{+89}_{-44}$ binary black hole (BBH) events and about $10^{+52}_{-10}$ binary neutron star (BNS) events over one year \cite{o4estimations}. Since neutron stars are more complex systems than black holes, they require more parameters to model and larger computational times. Specifically, the estimated time to analyze a BBH merger  event is on the order of days, while the estimated time to analyze a BNS merger event is on the order of weeks. As a result, the overall analysis time for the next run could be longer than one year. 
Second, with the addition of new observatories such as KAGRA and in the longer term the space interferometer LISA, there is hope to be able to observe different events such as supernovae explosions, the first candidate considered for detecting GWs, or massive black holes at galactic centers.

These highlights the importance of developing new algorithms that accelerate the analysis of these astrophysical events. In this manuscript, we review the most commonly used algorithms for the analysis of GWs and explore how quantum walk-based algorithms can be used to search for the set of parameters that best fits the data obtained in the detectors. In particular, we compare the performance of a quantum Metropolis-Hasting algorithm against its classical counterpart using the data from the events detected so far, which can be found in the PyCBC catalog at \cite{urlPYCBC, pycbcCATALOG}.

Quantum walks are a cornerstone family of quantum algorithms that exhibit a quadratically larger eigenvalue gap than the associated random walk. Since the inverse of the eigenvalue gap is often a multiplicative factor in the complexity of the algorithm, quantum walks are capable to outperform their classical counterparts, as we will explain. Specifically, quantum Metropolis-Hastings algorithms exhibiting this quadratic advantage have been developed~\cite{somma2007quantum,somma2008quantum,wocjan2008speedup,yung2012quantum,TTSpaper}, and have also been proposed as a method to tackle hard optimization problems~\cite{lemieux2021resource,QFold,QMS}. Although the use of quantum walks is quite extensive throughout the literature \cite{yasserpaper}, to the best of our knowledge there are no previous analysis of the use of this type of methods in the study of gravitational wave events, only in their detection~\cite{gao2022quantum}. Thus, we hope this work will be of interest to the gravitational astronomical community.

The rest of the manuscript is organized as follows. In the next section \ref{sec:algs} we introduce the basics of classical and quantum Metropolis-Hastings algorithms, and how they compare with each other. Then, in section~\ref{sec:simulation} we describe our simulation setup, whose results are explained in~\ref{sec:results}. We end the article with the conclusions and future work in section~\ref{sec:conclusions}.

\section{\label{sec:algs}Bayesian inference: classical and quantum Metropolis-Hastings algorithms}

The standard formalism for tackling the task of estimating the unknown parameters of a gravitational wave source classically is the formalism of Bayesian inference based on Markov Chain Monte Carlo methods (MCMC) \cite{LALinference}, as previously stated, despite the fact that several algorithms have been proposed in the literature. Among all the algorithms of this type, the Metropolis-Hastings algorithm stands out, and will consequently be the topic of study of this work, both in its classical and quantum versions.

\subsection{Classical Metropolis-Hastings}
The formalism of Bayesian inference describes the available knowledge about a parameter which we want to estimate from observations, $\theta$, as a probability density that we do not know a priori. In this framework, the Bayes' Theorem will be our main tool, where a prior probability distribution $p(\theta)$ is updated as new data $d$ is received from the experiment to give the posterior distribution $p(\theta | d)$ \cite{mcmcclasico2},
\begin{equation}\label{eqn:bayes}
    p(\theta | d)=\frac{p(\theta)p(d| \theta)}{p(d)}.
\end{equation}
The models we will explore have many parameters, in particular the models used for the BBH and BNS events contain 15 parameters \cite{pycbcCATALOG}, among which the distance to the source, the masses, the spin, the coalescence time or the position in the sky can be highlighted. We will indicate collectively this set of parameters as $\boldsymbol{\theta} = \{ \theta_1, \theta_2,...,\theta_N \}$. Thus, in order to describe the collective knowledge about all parameters and their relationships, the joint probability distribution on the multidimensional space $p(\boldsymbol{\theta} | d)$ will be used, being able to recover the probability for a specific parameter by marginalising over the other unwanted parameters.

Since the evidence, $p(d)$, only depends on the data obtained and is common for the whole parameter space, it can be taken as a normalisation factor ~\cite{mcmcclasico1}, so that the posterior probability reduces to
\begin{equation}\label{eqn:bayeslikeli}
    p(\boldsymbol{\theta} | d)\propto p(\boldsymbol{\theta}) \mathcal{L}(d| \boldsymbol{\theta}),
\end{equation}
where we have introduced the likelihood function as the conditional probability $\mathcal{L}(d| \bm{\theta}) := p(d| \bm{\theta})$. The likelihood function is something that we choose, it is a
description of the measurement. Thus, it is appropriate to choose a right likelihood function with the noise in our data. For this purpose, it is common in the literature of gravitational wave astronomy to assume a Gaussian noise in the detectors \cite{introbayes}, so that the Gaussian-noise likelihood function has the form:
\begin{equation}  \label{eqn:likelihood1}
    \mathcal{L}(d| \boldsymbol{\theta}) \propto \exp \left(-\frac{1}{2} \int_{0}^{\infty} \frac{| \tilde{d}(f)-\tilde{\mu}(\boldsymbol{\theta} , f) | ^2}{\sigma^2(f)}df \right),
\end{equation}
All these functions are computed in the Fourier domain, with $f$ denoting frequencies, where $\tilde{d}$ represents the data set obtained in the usual way at the detector, $\tilde{\mu}$ the waveform for a certain relativistic numerical model calculated from the parameter set $\boldsymbol{\theta}$, and $\sigma$ the detector noise (spectral density). It is interesting to note that despite having an integral at all possible frequencies, in practice we will limit this interval to those frequencies at which ground-based detectors work, also removing some frequencies such as the harmonics of domestic electricity, ensuring that at all these frequencies $\sigma$ is non-zero, and therefore there cannot be any divergence in (\ref{eqn:likelihood1}).

Since the likelihood function is equal to a negative exponential function, we can define an effective \textit{energy} function from the argument of that exponential, 
\begin{equation} \label{eqn:energy}
    E(\boldsymbol{\theta}) := \frac{1}{2} \int_{0}^{\infty} \frac{| \tilde{d}(f)-\tilde{\mu}(\boldsymbol{\theta} , f) | ^2}{\sigma^2(f)}df.
\end{equation}
This energy is a measure of how different is the available data, $\tilde{d}$, from our relativistic numerical model of gravitational wave, $\tilde{\mu}$, calculated from parameters $\boldsymbol{\theta}$. The use of numerical models of gravitational waveforms is due to the complexity of finding analytical solutions for some GW sources to the Einstein equations. For our purposes, we will use a single numerical model depending on the event (BBH or BNS). Then, the closer the parameters are to the source value, the closer the wave predicted by the relativistic numerical model resembles the data obtained, the lower the energy and therefore the higher the value of the likelihood.

The goal in the parameter estimation of GWs is to find the set of parameters that best fits the data obtained from the detectors. For this purpose, the Metropolis-Hastings algorithm is the standard and best-performing technique for parameter inference in gravitational wave astronomy, which performs a random walk over the gravitational wave model parameter space $\Omega$. 

This algorithm is specially designed to reach the equilibrium state as quickly as possible, that is, to reach the probability distribution $\phi$ such that $\mathcal{W}\phi = \phi$, where $\mathcal{W}$ is the transition matrix of our stochastic process. To reach this state we will start from an initial, usually uniform, distribution $\phi^{(0)}$ and a random transition will be proposed to obtain the next state. In this way:
\begin{enumerate}
    \item If at the current iteration $i$, the chain has a location in the parameter space of $\bm{\theta} _ i$, we propose a random transition $\Delta\bm{\theta }_ i$ to the new parameter state $\bm{\theta }_{i+1} = \bm{\theta} _i + \Delta \bm{\theta} _ i$.

    \item We now calculate the function \textit{likelihood} found in (\ref{eqn:likelihood1}) for the new state, and accept this transition with probability
    \begin{equation}
        \min \left(1,  \frac{p(\bm{\theta}_{i+1})}{p(\bm{\theta}_{i})} \left(\frac{\mathcal{L}(d| \bm{\theta}_{i+1})}{\mathcal{L}(d| \bm{\theta}_i)} \right)^{\frac{1}{T}}\right).
    \end{equation}
\end{enumerate}
The Metropolis-Hastings algorithm works by iterating over these steps. It is important to note that the variable $T$ introduced in the algorithm is what is known in the literature as \textit{tempering} \cite{tesisphd}, playing a role analogous to temperature, so that by varying this parameter we can obtain different speeds of convergence to the optimal parameters. We can go a step further and introduce simulated annealing, so we can start the optimisation with a high temperature and lowering it with some annealing scheme, $\beta$, so that at first the algorithm is more likely to accept larger steps and gradually accept smaller steps until finally converging to the optimal set parameters. 

Note that if the prior probability is uniform, we only need the function likelihood to decide whether to accept the next step in the random walk.  

We can finally formulate in matrix terms the random walk, and thus leave an algorithm ready for quantization. To do this, starting from a state $i$, the Metropolis-Hastings algorithm proposes a random change to one of the configurations, $j$, connected to $i$. We will call $T_{ij}$ the probability of such a proposal. If we have a uniform prior, it is customary to choose a uniform probability distribution,
\begin{equation} \label{eqn:Tij}
     T_{ij} = \begin{cases} \frac{1}{N} & \quad \text{there is a transition that connects \textit{i} with \textit{j}}, \\ 0& \quad \text{if there isn't},\end{cases} 
\end{equation}
for $N$ the number of possible transitions from state $i$. Alternatively, we could choose $T_{ij}\propto \frac{p(\bm{\theta}_{j})}{p(\bm{\theta}_i)}$.
Then, this change is accepted with a probability
\begin{equation}  \label{eqn:Aij}
    A_{ij} = \text{min} \left( 1, \left(\frac{\mathcal{L}(d| \bm{\theta}_{j})}{\mathcal{L}(d| \bm{\theta}_i)} \right)^{\beta}   \right)=\min \left(      1,e^{\beta [E(\bm{\theta}_j)-E(\bm{\theta}_i)]}    \right),
\end{equation}
resulting in an overall transition probability for a given step of $i \rightarrow j$:
\begin{equation}\label{eqn:Wij}
     W_{ij} = \begin{cases} T_{ij}A_{ij} & \quad \text{if } i\neq j,\\
     1-\sum_{k \neq j} T_{kj}A_{kj}& \quad \text{if } i = j .\end{cases} 
\end{equation}
As such, we have explained how the Metropolis-Hastings algorithm can be used to solve our Bayesian inference problem.

\subsection{Quantum Metropolis-Hastings}

Similarly, the objective for the Metropolis-Hastings algorithm is sampling the Boltzmann distribution $\rho^\beta = \ket{\phi^\beta} \bra{\phi^\beta} = Z^{-1} (\beta) \sum_{\theta \in \Omega} e ^{-\beta E(\bm{\theta})}  \ket{\bm{\theta}} \bra{\bm{\theta}} $. $E(\bm{\theta})$ is the energy function defined above for gravitational wave parameters $\bm{\theta}$, $\beta$ is the annealing schedule, playing the role of the inverse of the temperature and $Z (\beta) = \sum_{\theta \in \Omega} e ^{-\beta E (\bm{\theta})}$ is the partition function, a normalisation factor. Here, $\beta$ plays an important role in finding the lowest energy state because if it is large enough, only the configuration with the lowest possible energy will appear with a high probability when sampling the state.

\subsubsection{Construction of the Quantum Metropolis-Hastings Algorithm}

There are several ways of constructing quantum walks, but the best known for its results for this purpose is the one proposed by Szegedy~\cite{szegedy, portugal}. Szegedy's algorithm is based on two key features. First, it duplicates the registers encoding the state, so the operation space is twice the original Hilbert space.
Given the acceptance probabilities $\mathcal{W}_{ij} = T_{ij}A_{ij}$, defined previously in \eqref{eqn:Wij}, for the transition from state $i$ to state $j$, one defines the unitary operator
\begin{equation}\label{eqn:U_Szegedy}
    U\ket{j}\ket{0} := \ket{j} \sum_{i \in \Omega} \sqrt{\mathcal{W}_{ji}}\ket{i} = \ket{j}\ket{p_j}.
\end{equation}
The second key feature is that Szegedy quantum walk operator consists of two rotations similar to those performed in the well-known Grover's algorithm \cite{grovers,RMFgalindodelgado}.
Taking 
\begin{equation}
    R_0 := \mathbf{1} - 2\Pi_{0}= \mathbf{1} - 2(\mathbf{1}\otimes \ket{0}\bra{0})
\end{equation}
the reflection over the state $\ket{0}$ in the second Hilbert subspace, and $S$ the swap gate that exchanges both subspaces, we define the quantum walk step as 
\begin{equation}
 W:= U^\dagger S U R_0 U^\dagger S U R_0. \label{W definition}
\end{equation}
The first $R_0$ represents the first rotation, while $U^\dagger S U R_0 U^\dagger S U$ is the second.

The Metropolis-Hastings algorithm we are going to use is a modification of Szegedy's quantum walk \cite{TTSpaper}, replacing the bipartite space with the use of a coin. As a result, we will substitute the use of $W$ with a coin version of the evolution operator, $U_\mathcal{W}$. This operator will act on 3 quantum registers: a register for the system, which will indicate the current state of the system, a register encoding the transition, and a register for the Boltzmann coin encoding probabilities $A_{ij}$. We will denote them respectively by $\ket{x}_S \ket{m}_M \ket{b}_C$. The operator $U_\mathcal{W}$ is equivalent to half of the Szegedy operator $W$ in \eqref{eqn:Wij} under conjugation by the isomorphism operator $Y$, which maps the states in the second register of the original quantum walk to the transitions needed to reach them \cite{TTSpaper}:
\begin{equation}
    Y^{\dagger}(R_0 U^{\dagger} S U )Y = U_\mathcal{W}
\end{equation}
with 
\begin{equation}
    Y: \ket{x}_S \ket{y}_{S'} \mapsto \ket{x}_S \ket{m}_{M} \quad \text{such that } m(x) = y.
\end{equation}

The coined version operator $U_\mathcal{W}$ is then constructed as
\begin{equation} \label{eqn:EvOp}
    U_\mathcal{W} = R V^{\dagger} B^{\dagger} F B V.
\end{equation}
In this equation, $V$ prepares a superposition in the register $\ket{\cdot}_M$ over all possible transitions that the state may undergo in the next step. $B$ rotates the coin qubit $\ket{\cdot}_C$ to have an amplitude of $\ket{1}_C$ corresponding to the probability of acceptance $A_{ij}$ indicated in \eqref{eqn:Aij}. $F$ changes the register of the system $\ket{\cdot}_S$ to the new configuration, conditioned by the value of $\ket{\cdot}_M$ and $\ket{\cdot}_C$. Finally, $R$ is nothing more than a reflection on the $\ket{0}_{MC}$ state. If additional auxiliary registers $\ket{\cdot}_A$ are present, then the rotation $R$ will reflect on those registers too.

It is straightforward to construct these operators from the previous definitions. As $V$ prepares the possible transitions in the register $\ket{\cdot}_M$, these will be given by $T_{ij}$. The operator $V$ consequently prepares a uniform superposition
\begin{equation}
    V: \ket{0}_M \rightarrow N^{\sfrac{-1}{2}} \sum_m \ket{m}_M.
\end{equation}
Then $B$ prepares the coin qubit so that the probability of being in state $\ket{1}_C$ corresponds to the probability of acceptance, $A_{ij}$
\begin{equation}
\begin{split}
B &: \ket{x}_S \ket{m}_M \ket{0}_C \\
& \rightarrow \ket{x}_S \ket{m}_M \left( \sqrt{1-A_{x,m(x)}} \ket{0}_C +  \sqrt{A_{x,m(x)}}\ket{1}_C \right).
\end{split}
\end{equation}
$F$ will update the system registry to the new configuration,
\begin{equation}
\begin{split}
    F &: \ket{x}_S \ket{m}_M \ket{b}_C \\
    & \rightarrow \ket{b\cdot m(x)+(1-b)\cdot x}_S \ket{m}_M  \ket{b}_C .
    \end{split}
\end{equation}
And finally $R$ is the reflection operator on the state $\ket{0}_{MC}$
\begin{equation}
\begin{array}{cc}
     & R: \ket{0}_M \ket{0}_C \rightarrow - \ket{0}_M \ket{0}_C \\
     & \ket{j}_M \ket{b}_C \rightarrow \ket{j}_M \ket{b}_C \quad \text{if}\quad (j,b) \neq (0,0).
\end{array}
\end{equation}

There have been several proposals to quantise the Metropolis-Hastings algorithm using these quantum walks. These are often based on a slow evolution of $\ket{\theta_t} \rightarrow \ket{\theta_{t+1}}$ using quantum phase estimation to project the state into the stationary state of the corresponding Markov chain. Since the eigenvalue gap of the quantum walk operator is quadratically larger than the classical one, this technique often allows achieving the result with a quadratic quantum advantage~\cite{somma2007quantum,somma2008quantum,yung2012quantum,wocjan2008speedup}.
However, classically the Metropolis-Hastings algorithm is usually used with a rapidly varying annealing schedule, where no complexity-theoretical guarantees can be provided. 
As such, instead of relying on the complex quantum phase estimation procedure, we opt for a simpler and heuristic unitary procedure introduced first in Ref.~\cite{TTSpaper}. It consists of executing $L$ times the evolution operator on an initial state $U$
\begin{equation} \label{eqn:EvEstado}
    \ket{\psi (L)} := U_{\mathcal{W}^L} ...U_{\mathcal{W}^2} U_{\mathcal{W}^1} \ket{\phi^{(0)}}
\end{equation}
where $t=1,2,...,L$ will also define the annealing schedule, for the chosen values of $\beta(t)$ at each step. This is, in some ways, the simplest approach one could think of to quantize the Metropolis-Hastings algorithm, as it is very similar to the classical way of performing many steps of the random walk. \\

Thus, the procedure of our quantum algorithm is clear. Given an initial state $\ket{\phi^{(0)}}$ which represents a uniform superposition, the algorithm consists of applying the evolution operator defined in (\ref{eqn:EvOp}) iteratively, obtaining the state (\ref{eqn:EvEstado}). Hopefully, the probability of measuring the set of parameters that minimizes the energy, $\ket{\theta^*}$, will increase until it is close to 1. So, having introduced the quantum Metropolis-Hastings algorithm, we will next explore how to compare it with its classical counterpart.

\subsection{Figure of Merit: the Total Time to Solution}

\begin{figure*}[!htp]
  \includegraphics[width=1\textwidth]{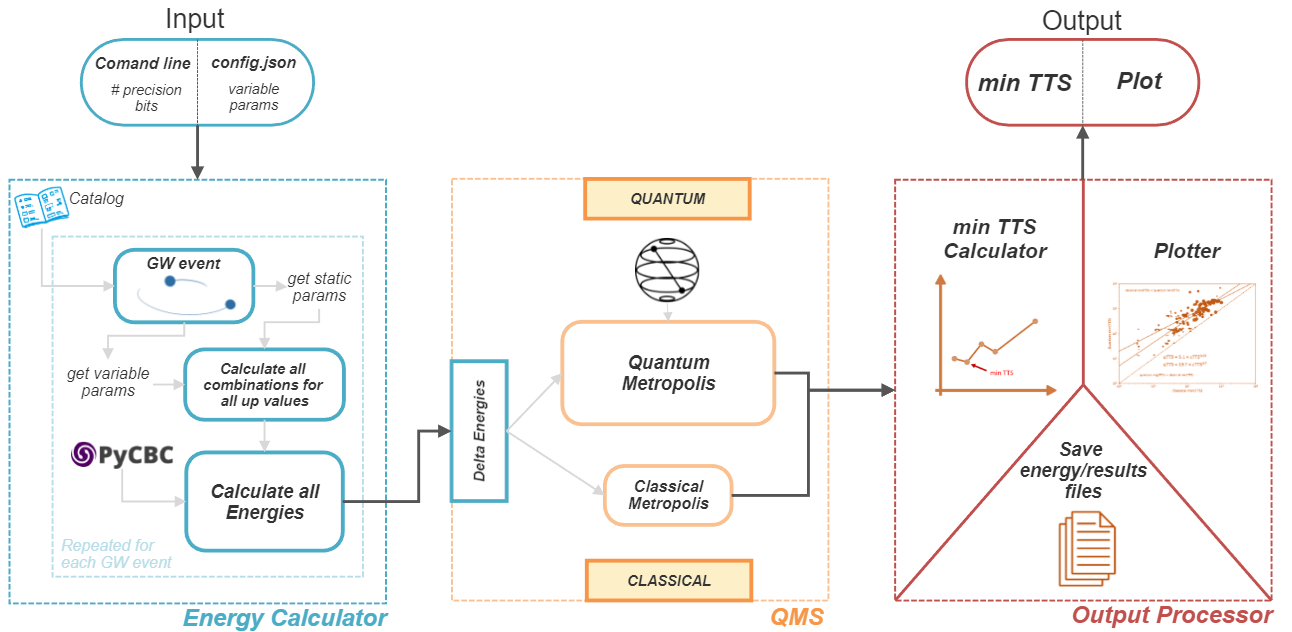}
  \centering
  \caption{Flowchart of the algorithm implemented. This algorithm consists of three stages: a first stage where, once the parameters to be inferred and their number of precision bits have been chosen, the energies are calculated with the help of the GW library PyCBC \cite{pycbc}. A second stage, where the two Metropolis-Hastings algorithms are executed, the quantum algorithm being programmed from the Qiskit library\cite{qiskit}, and a third stage where the results are stored and the values for the chosen figure of merit are calculated. Finally, the value of min(TTS) is obtained for each of the classical/quantum parts for later comparison.}
  \label{esquemaQGW}
\end{figure*}

Given the two algorithms, the classical algorithm and its quantum version, we are now interested in finding a tool capable of comparing both algorithms in a fair way. When we talk about a good algorithm for parameter inference in gravitational wave astronomy we are interested in two aspects: on the one hand, we want a model that finds the maximum likelihood parameters with a high probability. On the other, we want this algorithm to be fast. For example, computing all possible parameter configurations would be quite accurate, but extremely expensive. \\

When a random walk is employed to minimize some function $E(\bm{\theta})$, the minimum $\bm{\theta}^{*}$ is reached only with probability $p<1$. Starting from a certain distribution $q(\bm{\theta})$ and applying the $\mathcal{W}$ walk sequentially $t$ iterations, the probability of success is $p(t) = (\mathcal{W}^t q)(\bm{\theta}^{*})$. To increase this probability to a constant value of $1-\delta$, it is sufficient to repeat the procedure $L=\frac{\log(1-\delta)}{\log(1-p(t))}$ times. \\

Thus, we can find a natural metric given these conditions as the number of quantum walk steps $t$ times the number of attempts $L$, introducing the figure of merit called \textit{Total Time to Solution} (TTS) \cite{TTSpaper}, defined in other terms as the expected average time it would take for the algorithm to find the solution if we could repeat the procedure in case of failure:
\begin{equation}
    TTS(t)=t\frac{\log(1-\delta)}{\log(1-p(t))}
\end{equation}
where $t \in \mathbb{N}$ is the number of quantum/classical walk steps performed in one execution of the quantum/classical Metropolis-Hastings algorithm, $p(t)$ is the probability of hitting the correct parameter set after those iterations in each run of the algorithm, and $\delta$ is an arbitrary target probability of success, which we set to $0.9$. This metric represents the tradeoff between longer walks and the increased probability of success. Longer walks may achieve a higher probability of success and thus be repeated fewer times, but increasing the duration $t$ of the walk beyond a certain point could have a negligible impact on its probability of success $p(t)$ and may not be worth it. Thus we define the minimum \textit{Total Time to Solution} over the total of all iterations as $\min(\text{TTS}) = \min_t \text{TTS}(t) $, thus finding a way to compare quantum and classical walks by comparing the lowest achieved values of the TTS, i.e. comparing the $\min(\text{TTS})$ for different simulations. \\

It remains to explain how to calculate $p(t)$ in each of the algorithms. For the quantum case, the heuristic method of using the quantum walk explained in the previous section is proposed. Starting with a $\ket{\bm{\phi}^{(0)}}$ state, it consists of applying the quantum walk operator $U_{\mathcal{W}^j}$ sequentially, obtaining the state (\ref{eqn:EvEstado}). Thus, the probability of obtaining the searched state is:
\begin{equation}
    p(\bm{\theta} = \bm{\theta}^{*}) = |\bra{\bm{\theta}^*}U_{W^L}...U_{W^2}U_{W^1}\ket{\bm{\phi}^{(0)}}|^2
\end{equation}
So the construction of the TTS is straightforward, obtaining for the $L$-th iteration of the algorithm the expression
\begin{equation}
    TTS(L)=L\frac{\log(1-\delta)}{\log(1-|\bra{\bm{\theta}^*}U_{W^L}...U_{W^2}U_{W^1}\ket{\bm{\phi}^{(0)}}|^2)}.
\end{equation}
Similarly, for the classical algorithm, we estimate the probability of obtaining $\bm{\theta}^*$ after $t$ steps of the Metropolis-Hastings algorithm.

\section{\label{sec:simulation}Simulation Setup}

\begin{figure*}[!htp]
  \includegraphics[width=1\textwidth]{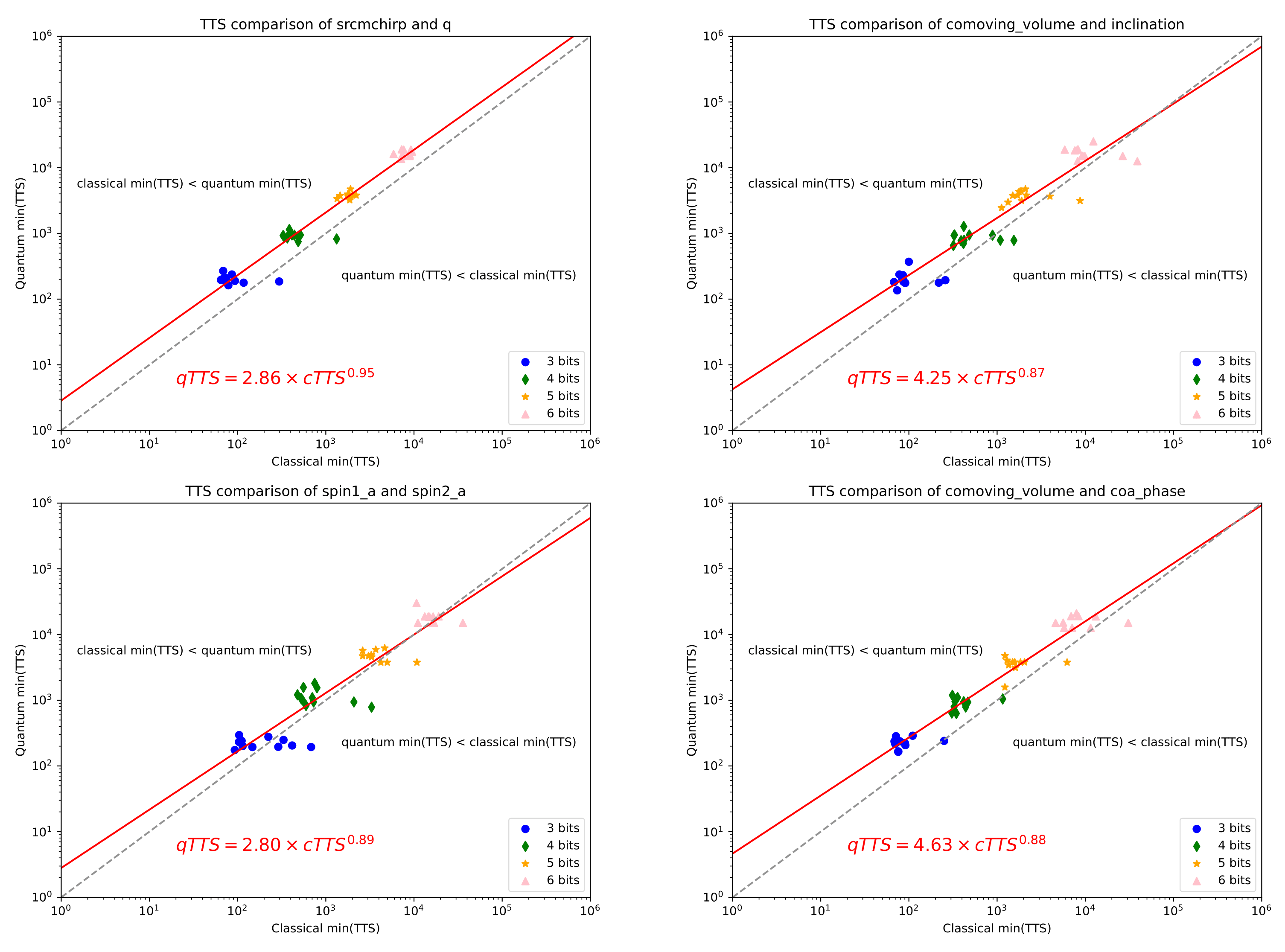}
  \centering
  \caption{Comparison of the minimum value of the classical and quantum TTS achieved for the 2-parameter inference simulations. Top left the source-frame chirp mass and the mass ratio have been inferred. Top right the comoving volume and the inclination have been inferred. Below left the dimensionless spin-magnitude of the larger and smaller object have been inferred. Below right the comoving volume and the coalescence phase have been inferred. All results have been obtained for 3-6 bits of precision with a constant value of $\beta$.}
  \label{2_params} 
\end{figure*}

In contrast to standard Bayesian inference analysis, the objective of this paper is to benchmark the usefulness of both classical and quantum Metropolis-Hastings algorithms more than to obtain the most likely parameter values.

To carry out this task, the fairest way to compare the two algorithms would be to run the classical Metropolis-Hastings algorithm on classical hardware and the quantum Metropolis-Hastings algorithm on quantum hardware. However, in the current era, only noisy intermediate-scale quantum computers are available, and these are only capable of running simple algorithms. Current quantum computers are not capable of executing the kind of algorithms proposed in this paper. This can be seen just by studying the cost of implementing the oracle (\ref{eqn:EvOp}). With the amount of qubits needed for each component of the Walk Operator \cite{lemieux2021resource} and the amount of parameters to be estimated with the desired precision, current quantum computers with a few hundred qubits would be unable to do it. Moreover, this has also been proven heuristically in \cite{QFold}. Under the same oracle construction as in this work, they have only managed to execute the calculation of two angles at the same time with only 1 bit accuracy and with a high amount of noise, so at this time only a simulation is achievable.

\begin{figure*}[!htp]
  \includegraphics[width=1\textwidth]{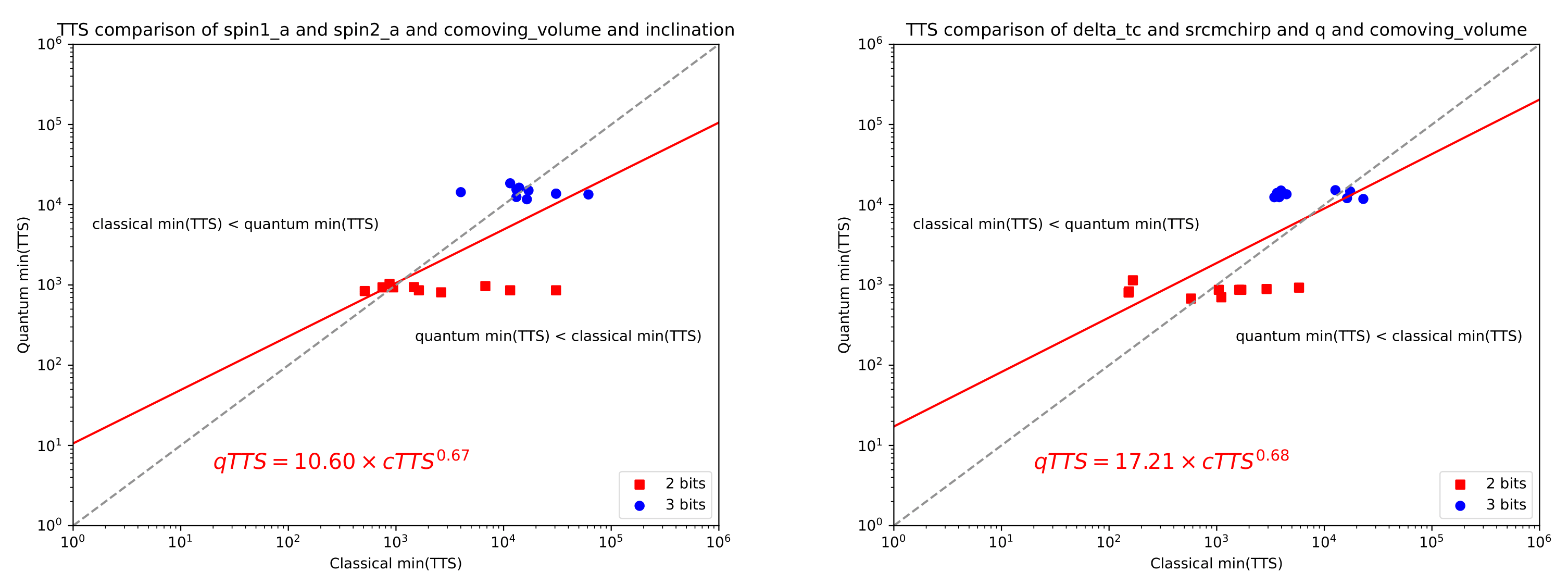}
  \centering
  \caption{Comparison of the minimum value of the classical and quantum TTS achieved for the 4-parameter inference simulations. On the left the dimensionless spin-magnitude of the larger and smaller object, the comoving volume and the inclination have been inferred. On the right the source-frame chirp mass, the mass ratio, the coalescence time and the comoving volume have been inferred. All results have been obtained for 2-3 bits of precision with a constant value of $\beta$.}
  \label{4_params} 
\end{figure*}

In this work, we have used the available data for GW mergers from the first GW catalog GWTC-1 \cite{gwtc-1}. The data and the posterior distributions inferred by \cite{pycbcCATALOG} can be found in ~\cite{urlPYCBC}. Since the quantum simulators are restricted in the number of qubits that we can simulate, we will fix most of the parameters to their inferred value, and only run the Metropolis-Hastings algorithm in a small subset of them. This limitation is due to the use of quantum simulators, that scale exponentially on classical computers, but it should not be an issue once we have early fault-tolerant quantum algorithms available. Likewise, the computation of the energy deltas going into the likelihood function is abstracted away to a readout table, which simplifies the quantum circuits required by our algorithm. While this requires exhaustively precomputing these values in advance, this is again a reflection of the limitations of current quantum simulators. When provided with a fault-tolerant quantum computer, any model to compute these energies can be implemented in a quantum computer at a constant multiplicative overhead. This is because quantum computers can perform classical calculations using exactly the same algorithms and therefore have the same scaling, differing only in a constant multiplicative overhead, which will be very high because an operation on a quantum computer takes 10 orders of magnitude longer than on a classical computer \cite{PRXQuantum.2.010103}. The quantum advantage does not happen in this energy calculation. Rather, the polynomial speedup associated with quantum walks will be reflected in the number of steps of the algorithm, the factor we measure with the Total Time to Solution introduced in the previous section.

The pipeline of our library is explained in figure \ref{esquemaQGW}, which we will explain in the following sections.

\subsection{The Energy Calculator Module}
The algorithm starts once we have chosen the set of parameters we want to infer from the configuration file and the number of precision bits with which we want to discretize the parameter space. For example, for a parameter such as the inclination, which takes values in the interval $[0, \pi]$, taking 1 bit of precision means that this parameter can only take the values $\{0, \pi\}$, while a precision of 2 bits indicates discretization to the values $\{0,\frac{\pi}{3},\frac{2\pi}{3}, \pi \}$.

Then, the real LIGO data of all GW events from GWTC-1 are downloaded. For each of these events, all possible combinations of the chosen parameters and number of precision bits are calculated and the energy of each of these combinations is stored. We have used the PyCBC gravitational wave library to obtain the energy of each combination using a model that marginalizes the polarization, in the same way as \cite{pycbc}, so we are left with a total of 14 parameters to vary for the BBH events listed in GWTC-1.

We further need the range of possible values for these parameters, which we extract from the configuration files of the repository of the catalog \cite{pycbcCATALOG}. Thus, this leaves a clear scheme for calculating the energy for any set of parameters for any BBH event for all observation periods.

\subsection{The Metropolis-Hastings Solver Module}

The Metropolis-Hastings Solver Module makes use of the software tool QMS \cite{QMS}. It receives a description of the problem as a list of tuples (state, energy) and it calculates the difference of energy between positions, called deltas. QMS tries, simultaneously, to find the minimum energy state and to avoid getting stuck in a local minimum energy state.

The main functionality of QMS is the quantum search to find the minimum energy state. However, it also serves as a test-bed for classical-quantum algorithm comparison. For this reason, QMS performs a classical search, in order to compare both results, classical and quantum. This software tool can return the resulting amplitude probabilities, as real quantum hardware does, or just sample from this target distribution.

QMS executes the quantum search in a quantum simulator running in a classical computer. The quantum simulator used in this work is \textit{QASM Simulator} of Qiskit \cite{qiskit}. Delta energies are introduced in the algorithm as a precalculated oracle, or table, queried via a serial QRAM. 

\begin{table*}[!htp]
\caption{\label{tab:fitexponents}Fitted exponents for different sets of variable parameters.}
\begin{tabular}{cc}
\multicolumn{2}{c}{2 PARAMETERS INFERENCE}                                    \\ \hline
\multicolumn{1}{|c|}{\textbf{Parameters inferred}} & \multicolumn{1}{c|}{\textbf{Fit Exponents}} \\ \hline
\multicolumn{1}{|c|}{Chirp mass and mass ratio}                  & \multicolumn{1}{c|}{0.95}             \\ \hline
\multicolumn{1}{|c|}{Dimensionless spin$_1$ and spin$_2$}                  & \multicolumn{1}{c|}{0.89}             \\ \hline
\multicolumn{1}{|c|}{Comoving volume and inclination}                  & \multicolumn{1}{c|}{0.87}             \\ \hline
\multicolumn{1}{|c|}{Comoving volume and coalesence phase}                  & \multicolumn{1}{c|}{0.88}             \\ \hline
\multicolumn{2}{c}{4 PARAMETERS INFERENCE}                                    \\ \hline
\multicolumn{1}{|c|}{\textbf{Parameters inferred}} & \multicolumn{1}{c|}{\textbf{Fit Exponents}} \\ \hline
\multicolumn{1}{|c|}{Coalesence time, mases and comoving volume}                  & \multicolumn{1}{c|}{0.68}            \\ \hline
\multicolumn{1}{|c|}{Dimensionless spin$_1$, spin$_2$, comoving volume and inclination}                  & \multicolumn{1}{c|}{0.67}            \\ \hline
\end{tabular}
\end{table*}

\subsection{The Output Processor Module}
Finally, the last module consists of generating all the results obtained by the Metropolis-Hastings algorithms for each of the GW events. As we have mentioned, our purpose is not to develop a quantum algorithm that estimates the parameters of the GW sources, but to demonstrate whether it is possible to achieve a quantum advantage in GW inference, thus creating a starting point for possible future developments. For this, we have previously introduced the figure of merit min(TTS). Such will be the output of this module of the algorithm. 

Thus, in this module the computation of the TTS will be carried out, storing the smallest value in TTS, the min(TTS), in order to later be able to study the behavior of this figure of merit as the complexity of the simulations increases. This complexity can be increased either by increasing the number of precision bits or by increasing the set of parameters to be inferred. 

We want to study the behavior of the classical min(TTS) versus the quantum min(TTS) as we increase the complexity of the simulation in order to find the exponent that will indicate the possible quantum advantage. In logarithmic scales, this reduces to finding the slope of the best linear least square fit.

There will be a quantum advantage when the quantum min(TTS) grows as a power of the classical min(TTS) with an exponent less than 1. For example, a quadratic quantum advantage implies that this exponent is $0.5$.

\section{\label{sec:results}Simulation Results}
In this section, we will test the classical and quantum Metropolis-Hastings algorithms. To achieve that goal we will study all the 11 events found in the first GW catalogue, the GWTC-1. In this catalog only BBH events have been detected \cite{gwtc-1}, so taking into account that we use the PyCBC model that marginalises the polarisation in order to reproduce the results of the catalogue of reference \cite{pycbcCATALOG}, we are left with 14 parameters available for the BBH events.

A good starting point is to make inference in pairs of parameters that make physical sense. Among these two-parameter sets we may highlight the two masses, reparameterised to chirp mass and mass ratio. Also the dimensionless spin-magnitude of the two black holes. In addition, and to complete the set of parameters, the luminosity distance, reparametrised to the comoving volume together with the inclination and the coalescense phase, are also two interesting pair of parameters.

As what we want to study is the slope, and this is obtained by increasing the complexity of the simulation, we will represent the previous pairs of parameters by increasing the number of precision bits. The results of these simulations will then be the plot of the quantum min(TTS) versus the classical one on a logarithmic scale to obtain in a simple and visual way the exponent. Starting with the inference of two parameters at a time, this is shown in Fig. \ref{2_params}.

The classical and quantum min(TTS) are on the x-axis and y-axis respectively on a logarithmic scale, so a slope less than one would imply an advantage of the quantum algorithms. Conversely, a slope greater than one would imply an advantage of the classical algorithms. To distinguish this behavior, in all these graphs there is a dashed gray line separating the space in which the quantum min(TTS) is less than the classical min(TTS) and vice versa. For the small sizes we have considered (low complexity) the quantum Metropolis provides modest results when compared to those obtained by the classical Metropolis. However, the key point of these plots is to note that the fact that the exponent is less than 1 leads us to expect that the quantum advantage dominates as the complexity increases and makes the quantum Metropolis more useful than its classical counterpart.

Furthermore, the next logical step to increase the complexity of the simulation is to infer more parameters at the same time. The results for this inference of 4 parameters  is shown in Fig. \ref{4_params}.

In summary, since we cannot run a quantum Metropolis-Hastings algorithm at high levels of complexity, we run the quantum and classical algorithms for simple cases and see how it behaves as the complexity increases. In this way, the fit for the calculation of the exponent is shown in Table~\ref{tab:fitexponents}.

Although only inferred on a small data set, this result may give us some hints as to whether our technique based on the use of a quantum Metropolis-Hastings algorithm, will be useful for estimating GW source parameters if a sufficiently large and error tolerant quantum computer is available, a point that will be discussed in the conclusions.

\section{\label{sec:conclusions}Conclusions and Outlook}
Throughout this work we have studied how quantum computation could complement classical techniques in the inference of gravitational wave parameters. To this aim, we have explored the feasibility of a Quantum Metropolis-Hastings Algorithm based on quantum walks for parameter inference in GW astronomy.

This algorithm has been built to run both the classical and quantum Metropolis-Hastings algorithm, since it is our aim to compare these two through the TTS figure of merit. For this purpose we have studied the first 11 gravitational wave events of the GWTC-1 catalogue.

It is currently not possible to run the Metropolis-Hastings quantum algorithms on real quantum hardware to compare them under the same conditions as their classical versions. However, it is possible to simulate the behavior of the quantum Metropolis-Hastings algorithm with classical hardware and compare their behaviors using an appropriate figure of merit, although we will effectively be limited in the size of the simulation, due to the number of complex numbers needed to describe a quantum system growing exponentially with the size of the system. 

Thus, we have studied the algorithms for simple but magnitude relevant examples in a black hole merger. Specifically, by inferring 2 and 4 event parameters at the same time, increasing the complexity of the simulation by increasing the number of precision bits to test the behavior of the quantum algorithm against the classical one.

In the case of inference of two parameters at the same time, it has been possible to run the algorithm up to 6 bits of precision, which implies a discretization comparable to the uncertainty of classical parameter estimation estimates \cite{pycbcCATALOG, gwtc-1}.

In the case of 4-parameter inference at the same time, the limit on the number of precision bits is reduced to half of the 2-parameter case above, due to the large computational cost of simulating a quantum processor on classical hardware. The maximum number of bits achieved in these simulations has been 3. It is important to note that this is not a limitation in the implementation of quantum Metropolis-Hastings algorithms to gravitational wave astronomy, it is only due to the large computational cost of simulating quantum algorithms on classical hardware, which will not be a problem when run on sufficiently large and fault tolerant quantum hardware.

In this way, in the 2-parameter inference a clear polynomial advantage has been obtained, achieving exponents of around $0.8 \sim 0.9$. This result is very interesting because, although a considerable number of points are already in the quantum advantage regime, this trend indicates that the advantage will become clearer the higher the complexity of the simulation.

Although we have dealt with problems of reduced sizes (complexity) where a polynomial quantum advantage is not yet useful, however the importance of our results is that they allow us to extrapolate this advantage to situations where the complexity of the simulation is very high, as it is in the inference of a larger set of parameters. The trend we found with this polynomial quantum advantage would imply an improvement of the quantum algorithms with respect to the classical ones, thus being able to reduce quantitatively and sensibly the execution times.

In the case of 4-parameter inference, the results indicate a higher polynomial advantage of quantum algorithms over classical algorithms than the 2-parameter case above. This could be due to the fact that we are increasing the complexity of the simulation by inferring more parameters at the same time. Still, it is interesting the fact that the vast majority of the points are already at a quantum advantage. Although it may seem that the number of events with 2 bits in the quantum advantage zone is higher than for 3 bits, which might lead one to believe that the quantum advantage is now smaller as the precision bits increase, the decisive factor that marks the possible quantum advantage while increasing complexity is the scaling exponent of the fitting. Then, we find that the scaling exponent is indeed smaller even than for the 2-parameter inference, finding an even more remarkable quantum advantage.

There are dozens of combinations in which to infer 2 parameters at the same time and hundreds in which to infer 4 parameters at the same time. There is also a high computational cost required for both the extraction of the energy files and the execution of the Metropolis-Hastings algorithms. Therefore, we wanted to perform the simulations on parameter sets where there is a good physical motivation to infer at the same time. The results obtained are very encouraging about the use of this type of quantum techniques in the study of gravitational radiation.

Thus, in this work we have proposed for the first time a quantum algorithm based on quantum walks in gravitational wave astronomy, achieving polynomial advantages over classical algorithms for the inference of a reduced set of physical parameters. This represents a first starting point for possible future developments and providing convincing arguments to motivate the use of quantum computation in the field of astrophysics.

\begin{acknowledgments}
We would like to thank Yifan Wang, Alex Nitz and Collin Capano on the usage of PyCBC. We acknowledge support from the CAM/FEDER Project No.S2018/TCS-4342 (QUITEMAD-CM), Spanish MINECO grants MINECO/FEDER Projects, PGC2018-099169-BI00 FIS2018, MCIN with funding from European Union NextGenerationEU (PRTR-C17.I1) and Ministry of Economic Affairs Quantum ENIA project. M. A. M.-D. has been partially supported by the U.S. Army Research Office through Grant No. W911NF-14-1-0103. P. A. M. C. thanks the support of a MECD grant FPU17/03620, and R.C. the support of a CAM grant IND2019/TIC17146.
\end{acknowledgments}

\newpage


\bibliography{arXiv_v1}

\end{document}